\documentclass[10pt,twocolumn,fleqn,aps,prl,groupedaddress]{revtex4}
\usepackage[latin9]{inputenc}
\usepackage{xcolor}
\usepackage{pdfcolmk}
\usepackage{float}
\usepackage{amsmath}
\usepackage{graphicx}
\PassOptionsToPackage{normalem}{ulem}
\usepackage{ulem}

\makeatletter

\providecolor{lyxadded}{rgb}{0,0,1}
\providecolor{lyxdeleted}{rgb}{1,0,0}
\DeclareRobustCommand{\lyxsout}[1]{\ifx\\#1\else\sout{#1}\fi}

\@ifundefined{textcolor}{}
{%
 \definecolor{BLACK}{gray}{0}
 \definecolor{WHITE}{gray}{1}
 \definecolor{RED}{rgb}{1,0,0}
 \definecolor{GREEN}{rgb}{0,1,0}
 \definecolor{BLUE}{rgb}{0,0,1}
 \definecolor{CYAN}{cmyk}{1,0,0,0}
 \definecolor{MAGENTA}{cmyk}{0,1,0,0}
 \definecolor{YELLOW}{cmyk}{0,0,1,0}
}

\pdfoutput=1
\usepackage{gensymb}
\usepackage{calrsfs}
\usepackage{float}
\usepackage{xcolor}
\usepackage{physics}

\makeatother

\begin{document}
\title{Stochastic Thermodynamics in a Non-Markovian Dynamical System}
\author{Cillian Cockrell$^{1}$ and Ian J. Ford$^{2}$}
\address{$^{1}$ School of Physics and Astronomy, Queen Mary University of
London, Mile End Road, London, E1 4NS, United Kingdom. \\
 $^{2}$ Department of Physics and Astronomy, University College London,
Gower Street, London, WC1E 6BT, United Kingdom.}
\begin{abstract}
The developing field of stochastic thermodynamics extends concepts
of macroscopic thermodynamics such as entropy production and work
to the microscopic level of individual trajectories taken by a system
through phase space. The scheme involves coupling the system to an
environment - typically a source of Markovian noise that affects the
dynamics of the system. Here we extend this framework to consider
a non-Markovian environment, one whose dynamics have memory and which
create additional correlations with the system variables, and illustrate
this with a selection of simple examples. Such an environment produces
a rich variety of behaviours. In particular, for a case of thermal
relaxation, the distributions of entropy produced under the non-Markovian
dynamics differ from the equivalent case of Markovian dynamics only
by a delay time. When a time-dependent external work protocol is turned
on, the system's correlations with the environment can either assist
or hinder its approach to equilibrium, and affect its production of
entropy, depending on the coupling strength between the system and
environment.
\end{abstract}
\maketitle

\section{Introduction}

The framework of thermodynamics derives immense power from its universality,
though initially this was limited only to macroscopic systems. Naturally
there has been much work to extend the framework to smaller systems
where fluctuations are non-negligible and where the power of universality
would be desirable. Nanoscale thermodynamics is a particularly important
area to explore because of the proximity to length scales where reversible
dynamical laws might more obviously apply; those from which irreversibility
is supposed to emerge. Various issues in this regard were raised by
Loschmidt and Zermelo over a century ago and are still discussed today.
Significant progress has been made in recent years towards resolving
these problems, starting with the work of Evans \textit{et al} \cite{evans93,evans95,evans02,evans04}
in deterministic thermostatted dynamics. The culmination of these
efforts to describe irreversibility is stochastic thermodynamics or
stochastic energetics \cite{seifertreview,sekimoto}. The treatment
of ``fast" environmental variables \cite{zwanzignoneq} provides
Markovian stochastic forces which are the source of irreversibility
in stochastic thermodynamics. In this view, the entropy production
is a stochastic, path-dependent property, based on the relative probability
of path reversal under the imposed stochastic and dissipative dynamics.
Stochastic entropy production has statistical properties which match
the properties of thermodynamic entropy, in particular that its expectation
value monotonically increases with time.

Here we consider a situation where we cannot regard an environment
as a set of fast variables in which we aren't interested. Without
this stratification, the environment's relaxation time can be large
enough to generate significant memory effects such that the system's
evolution becomes governed by a generalised Langevin equation (GLE)
\cite{zwanzignoneq}. We investigate this extension using the introduction
of an \textit{auxiliary system} lying between the system of interest
and a Markovian bath. The auxiliary system can be interpreted a mathematical
device to generate the memory effects arising from environments that
do not relax sufficiently quickly. There are numerous advantages to
this approach: the memory effects can be encoded into simple evolution
equations of the auxiliary system which avoids the complications of
analysing or simulating non-Markovian noise, and since the system
and auxiliary system combined constitute a single dynamical entity
that couples to a Markovian bath, the total entropy production, heat
transfer and work all obey the established fluctuation theorems of
standard Markovian stochastic thermodynamics. It's important to note
that, so long as it can generate the desired memory effects, this
``globally" Markovian set-up is a perfectly valid representation
of a non-Markovian environment.

Non-Markovian dynamics and GLEs have received much attention in non-equilibrium
molecular dynamics and solid state physics \cite{stella14,ceriotti09,morrone11,kantorovich08},
nano-scale quantum thermodynamics \cite{kutvonen15,strasberg16,kato16,wang15,nicolin11},
biophysics \cite{garg85,fulinski98}, and from a general theoretical
point of view \cite{speck07,seifert16,campisi09,jarzynski04,kawamoto11,ohkuma07,anders17}.
The usage of auxiliary systems (or reaction coordinates) as an efficient
algorithm to generate non-Markovian dynamics has been independently
developed and implemented multiple times for both classical and quantum
systems \cite{stella14,strasberg16,garg85,martinazzo11,woods14,chin10}.
In particular, the work of Strasberg \cite{strasberg16} contains
similar observations to ours, though in the context of quantum rather
than classical systems, and with an emphasis on work and efficiency
rather than entropy production and stochastic thermodynamics.

In Section \ref{sec:Generation-of-Coloured} we introduce the approach
using a simple elaboration of the Ornstein-Uhlenbeck process. We discuss
the interpretation of the auxiliary system in Section \ref{sec:Interpretation-of-the}.
We identify how to approach the Markovian limit of a non-Markovian
environment in Section \ref{sec:The-Markovian-Limit}. Then we present
a series of examples, starting in Section \ref{sec:A-Statically-Tethered}
with a harmonically tethered Brownian particle undergoing thermal
relaxation when coupled to a non-Markovian environment. We find that
non-Markovian environments cause the production of the same amount
of entropy, only more spread out in time. The next example in Section
\ref{sec:A-Dynamically-Tethered} involves a time-dependent external
protocol of work performed on the Brownian particle, and we find there
can be either an increase or reduction in the entropy production and
heat flows, depending on the relative time scales of change in the
protocol and the auxiliary system. Our last example, in Section \ref{sec:A-Driven-System},
concerns an untethered particle subjected to periodic driving, while
coupled to an environment, to demonstrate further features of thermodynamic
response for non-Markovian noise. Our conclusions are given in Section
\ref{sec:Conclusions}.

\begin{figure}
\begin{centering}
{\scalebox{0.28}{\includegraphics{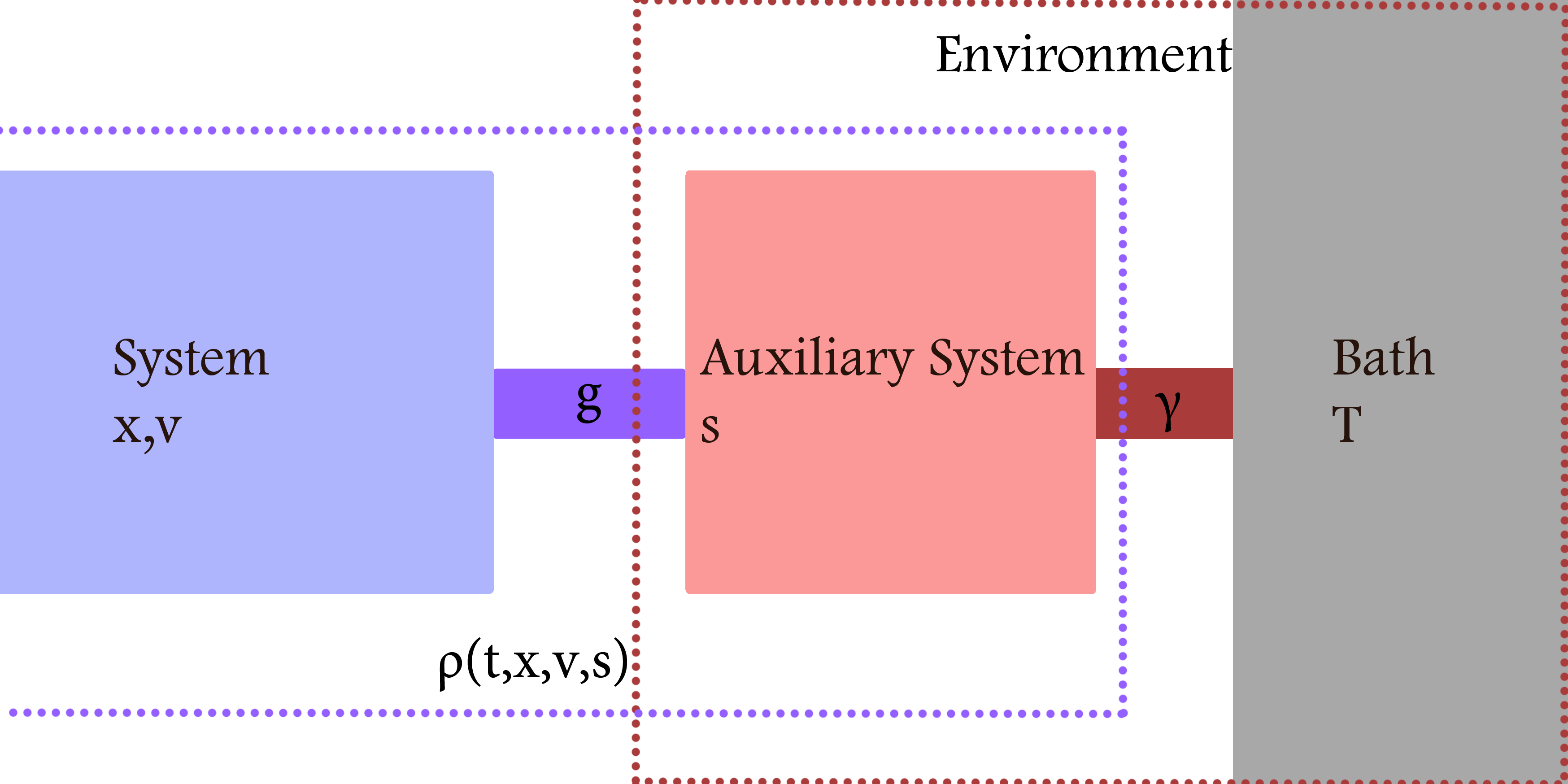}}}
\par\end{centering}
\caption{Schematic diagram of a system that receives the dissipative and stochastic
influence of the Markovian bath \textit{via} the auxiliary system
interposed between them. Note the discrimination between the term
\textbf{environment} (generic term for those degrees of freedom of
the world that we choose not to perceive) and \textbf{bath} (an entity
consisting of a large number of degrees of freedom with an ``infinitely
fast" relaxation time). $\rho$ is a time-dependent probability density
function over the dynamical variables of the system and auxiliary
system, and $g$ and $\gamma$ are coupling constants.}
\label{diagram}
\end{figure}

\section{Generation of Coloured Noise from White Noise\label{sec:Generation-of-Coloured}}

The use of white noise makes it simple to derive an expression for
the entropy production so we are therefore strongly incentivised to
represent non-Markovian dynamics using white noise. To illustrate
such a procedure, we consider a one dimensional Ornstein-Uhlenbeck
process with unit mass: 
\begin{equation}
\dv{v(t)}{t}=-\gamma v(t)+\sqrt{2\gamma T}\xi(t),\label{ornsteinuhlenbeck}
\end{equation}
with velocity $v$, friction coefficient $\gamma$ and $\xi(t)$ the
usual Gaussian white Markovian noise provided by a heat bath. We have
absorbed the Boltzmann constant $k_{B}$ into the heat bath temperature
$T$ for convenience. We now introduce revised dynamics featuring
a new variable, $s(t)$ (describing the \textit{auxiliary system}),
lying dynamically ``between" the system and the bath: 
\begin{equation}
\dv{v(t)}{t}=gs(t),\label{dvdt}
\end{equation}
\begin{equation}
\dv{s(t)}{t}=-\gamma s(t)-gv(t)+\sqrt{2\gamma T}\xi(t),\label{dsdt}
\end{equation}
where $g$ is a coupling constant. The situation is illustrated in
Fig. \ref{diagram}, for a system with a position coordinate $x$
as well as a velocity. We can solve equation (\ref{dsdt}) using the
integrating factor $\mathrm{e}^{\gamma t}$: 
\begin{equation}
s(t)=\int_{-\infty}^{t}\dd t^{\prime}\left(-gv(t^{\prime})+\sqrt{2\gamma T}\xi(t^{\prime})\right)\mathrm{e}^{-\gamma(t-t^{\prime})},\label{s(t)}
\end{equation}
and substitute this into equation (\ref{dvdt}), giving 
\begin{align}
\begin{split}\dv{v(t)}{t}= & -g^{2}\int_{-\infty}^{t}\dd t^{\prime}\ v(t^{\prime})\ \mathrm{e}^{-\gamma(t-t^{\prime})}\\
 & +g\sqrt{2\gamma T}\int_{-\infty}^{t}\dd t^{\prime}\ \xi(t^{\prime})\ \mathrm{e}^{-\gamma(t-t^{\prime})}.
\end{split}
\label{dvdtsub}
\end{align}
Equation (\ref{dvdtsub}) resembles equation (\ref{ornsteinuhlenbeck}),
except with memory kernels in the friction and noise terms (\textit{i.e.}
it is a generalised Langevin equation). It is a non-Markovian equivalent
of the Ornstein-Uhlenbeck process.

The situation is as follows: $v$ is a measurable system parameter
(a velocity), and the auxiliary system variable, $s$, is part of
the environment, which in this case iscomprised of the auxiliary system
and the heat bath which supplies the white noise. We postulate that
the variable $s$ has even time-reversal symmetry. The constant $g$
specifies the coupling strength between the particle's velocity and
the environment, and we note that the particle only ``sees" the
bath indirectly through the auxiliary system. Because $v$ appears
in the evolution equation (\ref{dsdt}) of $s$, the dynamics of the
environment are correlated with those of the system. This feature
is responsible for the non-Markovian noise and the interesting thermodynamic
consequences that this noise causes.

\section{Interpretation of the Auxiliary System\label{sec:Interpretation-of-the}}

The degree of freedom $s$ has been designated to be part of the environment,
which suggests that it should be considered unmeasurable, much like
the huge number of degrees of freedom which make up the heat bath
to which it is connected. The utility of baths is that their entropy
production can be quantified without needing to measure their microscopic
degrees of freedom.

But for a non-Markovian environment this is not true; its correlations
with the system mean its behaviour cannot be characterised just by
its temperature. This is what makes it more challenging to consider
within stochastic thermodynamics, and, of course, all physical environments
are to some degree non-Markovian. But we are not obliged to treat
the auxiliary system as a physical element of the environment: we
can instead simply regard it as a tool that is used in conjunction
with the white noise in order to generate a desired memory effect
in our system (\textit{i.e.} equation (\ref{dvdtsub})). By careful
use of auxiliary systems, one can imagine the construction of intricate
memory kernels while still having a world that, as a whole, is represented
by a Markovian bath (or baths) connected to some observable degrees
of freedom and thus open to analysis by stochastic thermodynamics.

Some reflection confirms that we are not working at a different level
of coarse graining when we introduce an auxiliary system. Varying
the level of coarse graining can change the degree of production of
entropy in a given process since entropy is dependent on uncertainty,
and reduces when we better monitor the evolution of the world. We
are merely reformulating the non-Markovian stochastic equations of
motion of the system in a fashion that allows standard thermodynamic
analysis. The auxiliary system coordinate is a function of system
dynamical variables and a noise, as illustrated in equation (\ref{s(t)}).
It is \emph{not} an unobservable parameter of a physical bath. The
entropy production arising from the reformulated dynamics is equivalent
to the desired entropy production associated with the non-Markovian
stochastic dynamics.

\section{The Markovian Limit\label{sec:The-Markovian-Limit}}

The superficial similarities between equations (\ref{ornsteinuhlenbeck})
and (\ref{dvdtsub}) led us to regard equation (\ref{dvdtsub}) as
the non-Markovian equivalent of equation (\ref{ornsteinuhlenbeck}).
We now establish their detailed relationship. We will use the (\emph{non}-)\emph{Markovian
case} as a shorthand to refer to a system connected to a (non-)Markovian
environment.

Setting the coupling constant, $\gamma$, between the auxiliary system
and the bath very large allows the integrals in equation (\ref{dvdtsub})
to be approximated, but the relationship between the fluctuation and
dissipation coefficients in equation (\ref{dvdtsub}) doesn't quite
match that in (\ref{ornsteinuhlenbeck}). To formulate how the coupling
constants should behave in order to maintain this relationship, we
consider what must happen to the auxiliary system in order for the
environment to be considered Markovian. We assert that in the Markovian
limit of our models, $\dd s/\dd t$ must vanish. The best justification
for this behaviour is that if $\gamma$ becomes very large, the coupling
between the auxiliary system and the bath becomes strong enough that
any disturbance in $s$ brought about by the system is immediately
dissipated.

However the equivalent Markovian case would not be recovered in the
limit $\gamma\rightarrow\infty$ alone, even though this limit offers
Markovian dynamics of some sort. The reason for this is as follows.
As $\gamma$ increases, the auxiliary system effectively loses its
dynamical freedom and never appreciably departs from equilibrium with
the bath, no matter the state of the system. However the system is
not coupled directly to the bath but just to an auxiliary system that
is in a very rigid equilibrium. In a sense the system is attached
to a microscopic Markovian bath which consists of one degree of freedom,
$s$.

\begin{figure}
\begin{centering}
{\scalebox{0.5}{\includegraphics{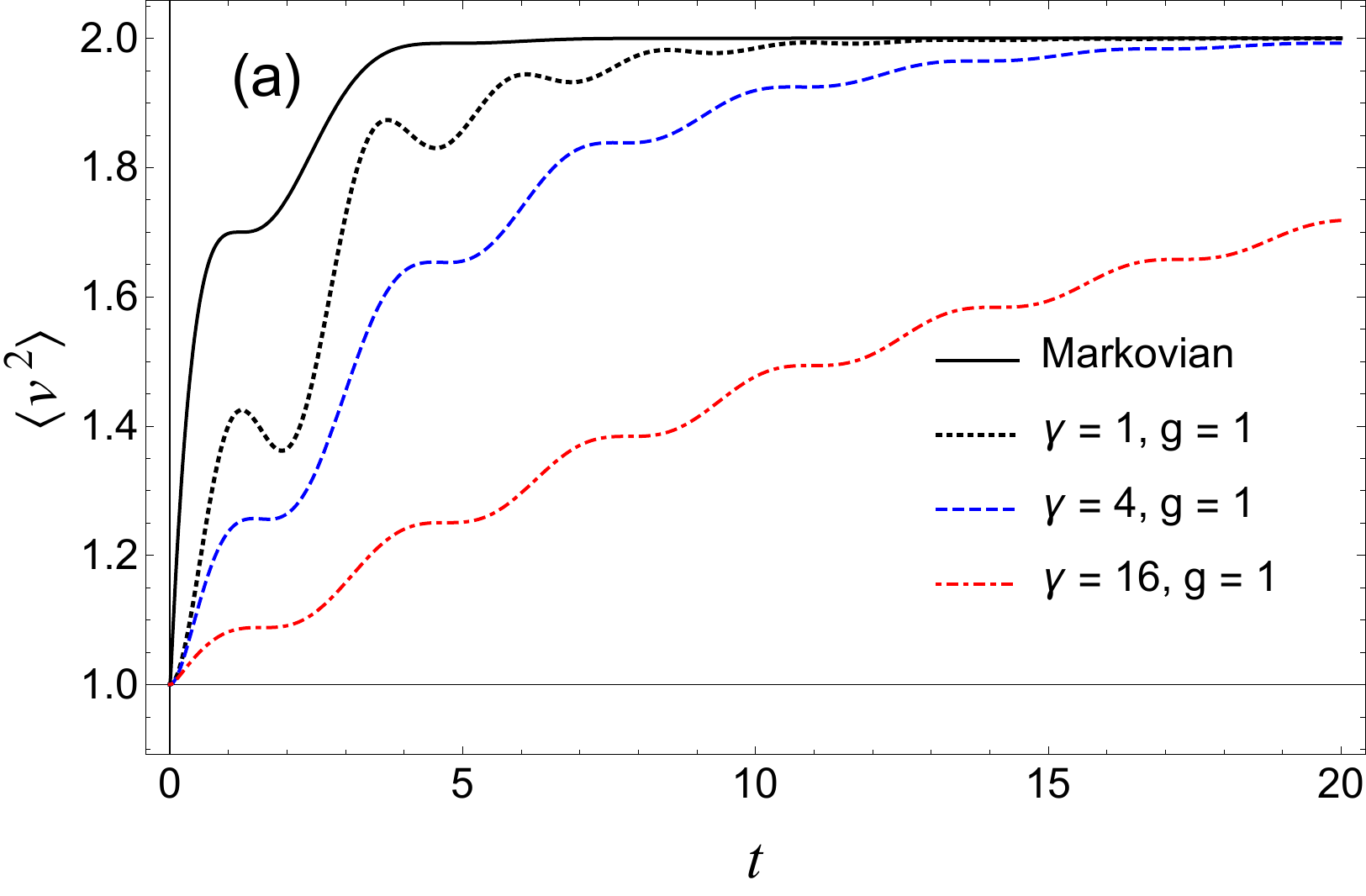}}} {\scalebox{0.5}{\includegraphics{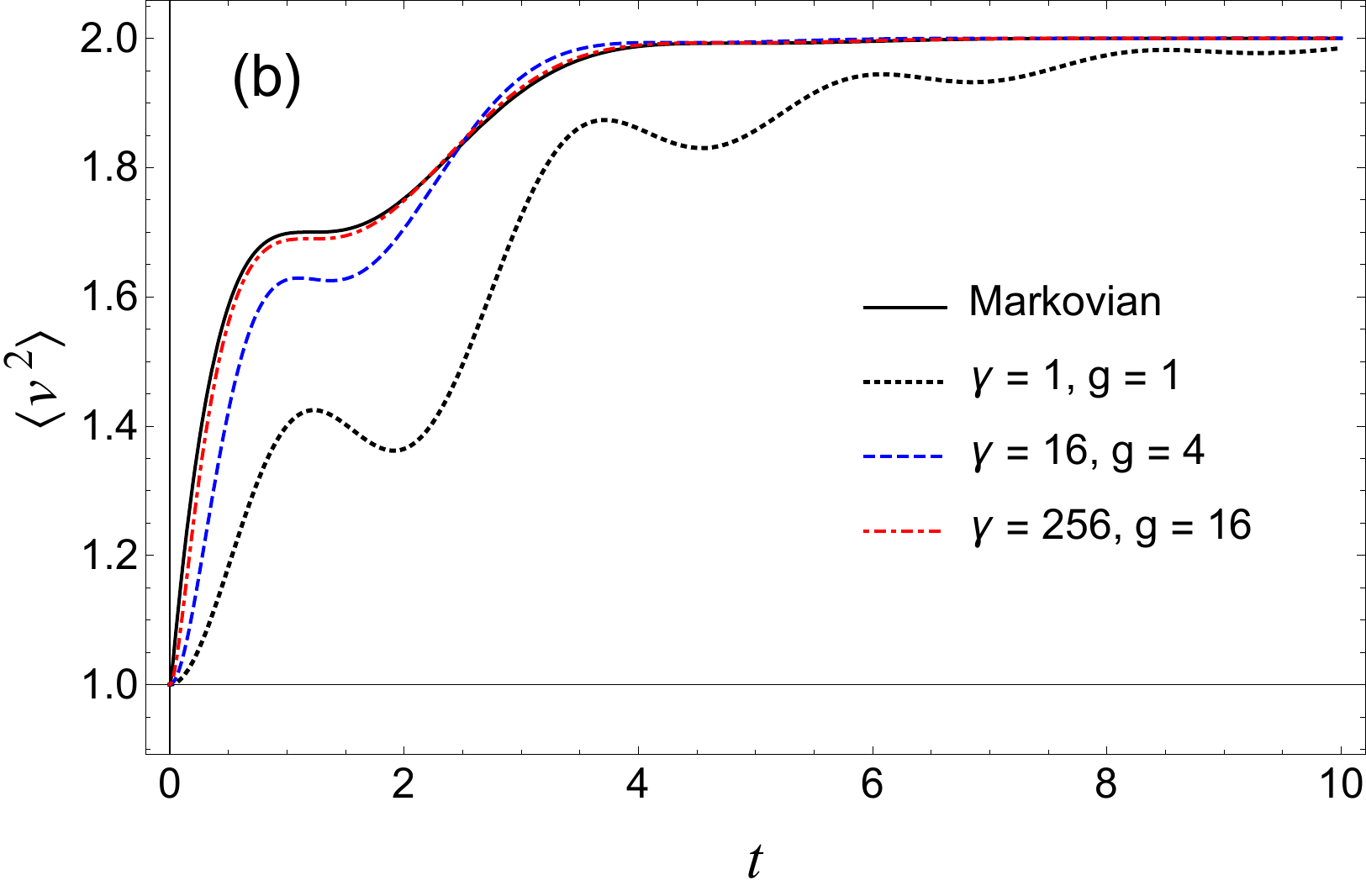}}}
{\scalebox{0.5}{\includegraphics{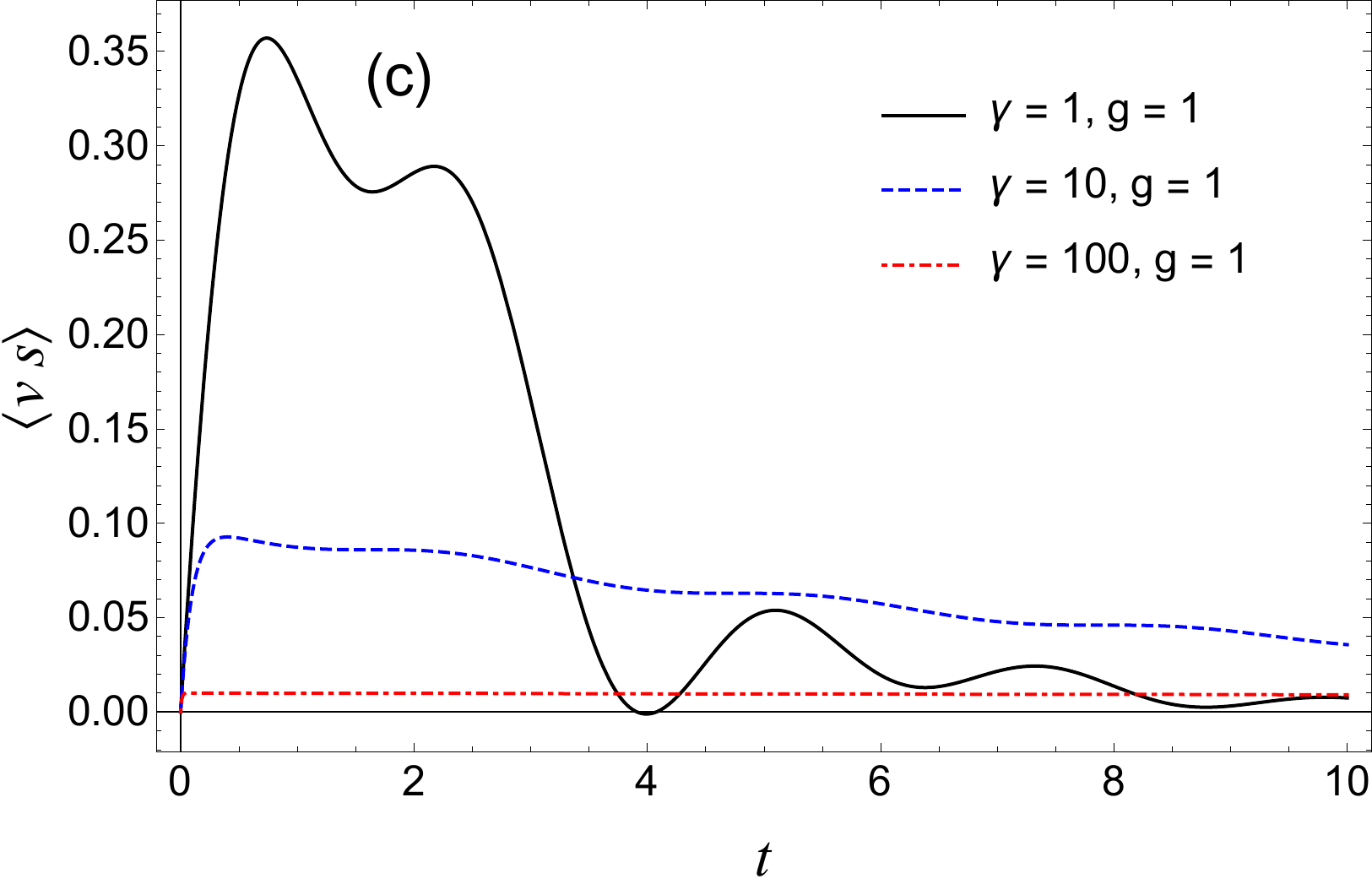}}}
\par\end{centering}
\caption{Evolution of $\langle v^{2}\rangle$ according to equation (\ref{dvdtsub})
for an instantaneous change in temperature from $1$ to $2$ at $t=0$,
starting at equilibrium, with various auxiliary system-bath coupling
strengths with (a): the system-auxiliary system coupling $g$ kept
constant, (b): the ratio $g^{2}/\gamma$ maintained at unity; and
with the Markovian evolution of $\langle v^{2}\rangle$ according
to equation (\ref{ornsteinuhlenbeck}) at $\gamma=1$ for reference.
(c): Evolution of the system-auxiliary system correlation $\langle vs\rangle$
for a range of auxiliary system-bath coupling strength $\gamma$ with
the system-auxiliary system coupling $g$ kept constant. .}
\label{markovianlimit}
\end{figure}

We explore this in Fig. \ref{markovianlimit} where we study the dynamics
of relaxation of the system velocity $v$ in response to an increase
of the temperature $T$ of the bath for a variety of coupling strengths
$g$ and $\gamma$. A microscopic bath at large $\gamma$ but finite
$g$ would not relax the system to equilibrium as quickly as a ``genuine"
(macroscopic) bath in the absence of an auxiliary system, which is
exactly the behaviour seen in Fig. \ref{markovianlimit} (a). However
the dynamics do still become Markovian, as evidenced by the vanishing
correlation between $v$ and $s$ in Fig. \ref{markovianlimit} (c).
By increasing the coupling between the system and auxiliary system
with the heat bath, $g$, while adopting a large $\gamma$ to preserve
the auxiliary system's equilibrium, we can allow the auxiliary system
to transfer enough heat to or from the system to resemble a macroscopic
Markovian bath with coupling $\tilde{\gamma}$. The relationship between
these parameters turns out to be $g^{2}=\gamma/\tilde{\gamma}$ (see
Fig. \ref{markovianlimit} (b)).

We can motivate this ratio using equation (\ref{dsdt}). The dissipation
from the bath should instantaneously disperse any fluctuations in
the auxiliary system coordinate: 
\begin{equation}
\dv{s}{t}=-\gamma s-gv+\sqrt{2\gamma T}\xi\approx0.
\end{equation}
We multiply through by $g/\gamma$ and rearrange:
\begin{equation}
gs=-\frac{g^{2}}{\gamma}v+\sqrt{\frac{2g^{2}T}{\gamma}}\xi,
\end{equation}
and substitute into equation (\ref{dvdt}) to give
\begin{equation}
\dv{v}{t}=-\tilde{\gamma}v+\sqrt{2\tilde{\gamma}T}\xi,
\end{equation}
with $\tilde{\gamma}=g^{2}/\gamma$. A non-Markovian environment can
therefore be tuned smoothly to behave like a Markovian bath. We refer
to limit of $g,\gamma\rightarrow\infty$ while holding $g^{2}/\gamma$
constant as the \textit{Markovian limit} of the non-Markovian dynamics
We note that since this limit reduces the relaxation time of $s$
to that of the bath, the constants $\gamma$ and $g$ are implicitly
a measure of the auxiliary system's relaxation time.

\section{Relaxation of a Tethered Particle\label{sec:A-Statically-Tethered}}

We now consider a tethered harmonic oscillator of unit mass coupled
to a non-Markovian environment. The spring constant $\kappa$ is time
independent. The Hamiltonian for this system is: 
\begin{equation}
\mathcal{H}(x,v)=\frac{1}{2}\kappa x^{2}+\frac{1}{2}v^{2},\label{tetheredhamiltonian}
\end{equation}
where $x$ is the displacement. This allows us to specify the equilibrium
statistics of the microscopic variables (irrespective of coupling
strengths): 
\begin{equation}
\frac{1}{2}\kappa\langle x^{2}\rangle=\frac{1}{2}T,\label{varx}
\end{equation}
and similarly for $v$. We postulate that $s$ shares this equilibrium
relationship, \textit{i.e.}
\begin{equation}
\frac{1}{2}\langle s^{2}\rangle=\frac{1}{2}T.\label{vars}
\end{equation}
This gives the equilibrium probability density function (pdf) of the
system plus auxiliary system a Gaussian form 
\begin{equation}
\rho_{{\rm eq}}^{\mathrm{NM}}(x,v,s,T)=\frac{\sqrt{\kappa}}{(2\pi T)^{\frac{3}{2}}}\exp\left(-\frac{\kappa x^{2}}{2T}-\frac{v^{2}}{2T}-\frac{s^{2}}{2T}\right).\label{tetheredcanonical}
\end{equation}
We employ the ansatz that the non-equilibrium probability density
maintains the Gaussian form, and write: 
\begin{align}
\begin{split}\rho^{\mathrm{NM}}(t,x,v,s)= & \exp\left(N-Ax^{2}-Bv^{2}-Cs^{2}\right.\\
 & \left.-Dxv-Exs-Fvs\right).
\end{split}
\label{tetheredpdf}
\end{align}
where the coefficients $N$ and $A$ through $F$ are all functions
of time $t$.

The stochastic entropy production then has incremental contributions
from the system, auxiliary system and the bath \cite{seifertreview}:

\begin{align}
\begin{split}\dd\Delta s_{\mathrm{tot}} & =-\dd\ln\rho^{\mathrm{NM}}(t,x,v,s)+\dd\Delta s_{\mathrm{bath}}\\
 & =-\dd\ln\rho^{\mathrm{NM}}(t,x,v,s)-\frac{\dd Q}{T},
\end{split}
\label{entropyequation}
\end{align}
where $\dd Q$ is the incremental heat transferred to the system.
We write the equations of motion in It\^{o} form allowing us to use the
expression for $\dd\Delta s_{\mathrm{bath}}$ from ref. \cite{spinneyfordpre}.
For simplicity, we set $m=1$ and find
\begin{equation}
\dd x=v\dd t,\label{tetheredx}
\end{equation}
\begin{equation}
\dd v=-\kappa x\dd t+gs\dd t,\label{tetheredv}
\end{equation}
\begin{equation}
\dd s=-\gamma s\dd t-gv\dd t+\sqrt{2\gamma T}\dd W,\label{tethereds}
\end{equation}
with $\dd W$ an increment of the Wiener process. The total entropy
production (see the appendix for derivation) is then

\begin{widetext} 
\begin{align}
\begin{split}\dd\Delta s_{\mathrm{tot}}^{\mathrm{NM}}= & -\dd t\left(2\gamma-4\gamma TC+\gamma TE^{2}x^{2}+\gamma TF^{2}v^{2}+\left(4\gamma TC^{2}-\frac{\gamma}{T}\right)s^{2}+2\gamma TEFxv+4\gamma TCFvs\right)\\
 & +\dd W\sqrt{2\gamma T}\left(Ex+Fv+\left(2C-\frac{1}{T}\right)s\right),
\end{split}
\label{tetheredentropy}
\end{align}
where $N$ and $A$ to $F$ are found by manipulating the Fokker-Planck
equation
\begin{align}
\begin{split}\pdv{\rho^{\mathrm{NM}}}{t}= & -\pdv{v\rho^{\mathrm{NM}}}{x}-\pdv{(-\kappa x+gs)\rho^{\mathrm{NM}}}{v}-\pdv{(-\gamma s-gv)\rho^{\mathrm{NM}}}{s}\\
 & +\gamma T\pdv[2]{\rho^{\mathrm{NM}}}{s},
\end{split}
\label{tetheredfpe}
\end{align}
as shown in the appendix.

Similarly, the pdf for the Markovian dynamics is assumed to take the
form

\begin{align}
\begin{split}\rho^{\mathrm{M}}(t,x,v)= & \exp\left(\tilde{N}-\tilde{A}x^{2}-\tilde{B}v^{2}-\tilde{C}xv\right).\end{split}
\label{markoviantetheredpdf}
\end{align}
 The It\^{o} equations of motion in the Markovian case are 
\begin{equation}
\dd x=v\dd t,\label{markoviantetheredx}
\end{equation}
\begin{equation}
\dd v=-\kappa x\dd t-\tilde{\gamma}v\dd t + \sqrt{2\tilde{\gamma}T}\dd W,\label{markoviantetheredv}
\end{equation}
and the increment in total entropy production is 
\begin{align}
\begin{split}\dd\Delta s_{\mathrm{tot}}^{\mathrm{M}} & =-\dd t\left(2\tilde{\gamma}-4\tilde{\gamma}T\tilde{B}+\tilde{\gamma}T\tilde{C}^{2}x^{2}+\left(4\tilde{\gamma}T\tilde{B}^{2}-\frac{\tilde{\gamma}}{T}\right)v^{2}+4\tilde{\gamma}T\tilde{C}xv\right)\\
 & +\dd W\sqrt{2\tilde{\gamma}T}\left(\tilde{C}x+\left(2\tilde{B}-\frac{1}{T}\right)v\right),
\end{split}
\label{markoviantetheredentropy}
\end{align}
with time dependent functions $\tilde{N}$, $\tilde{A}$, $\tilde{B}$
and $\tilde{C}$ as shown in the appendix.

\end{widetext}The total entropy production along a given trajectory
for both cases with an isothermal bath is given by integrating equation
(\ref{entropyequation}):

\begin{equation}
\Delta s_{\mathrm{tot}}=-\left(\Delta\ln\rho+\frac{Q}{T}\right).\label{integratedentropy}
\end{equation}
We consider in this case an instantaneous change in the temperature
of the \textit{environment} (\emph{i.e.} the auxiliary system plus
the bath) from $T_{0}$ to $T$. The system in equilibrium at temperature
$T_{0}$ is brought into contact with the environment in equilibrium
at temperature $T$, for both the Markovian and non-Markovian environments.
Because the auxiliary system is part of the environment, the process
begins and ends with it in thermal equilibrium at temperature $T$,
therefore it contributes nothing to the term $\Delta\ln\rho$ once
it has regained equilibrium. The term $Q/T$ therefore determines
the difference between the equilibrium to equilibrium entropy productions
for the Markovian and non-Markovian cases. The system plus auxiliary
system is closed apart from its thermal contact with the bath and
therefore the heat transfer exactly equals its energy change. Again,
since the auxiliary system starts in thermal equilibrium with the
bath, the mean system energy change in the non-Markovian case is the
same as that in the Markovian case. Differences inmean entropy production
between the Markovian and non-Markovian cases can therefore appear
only during non-equilibrium behaviour.

\begin{figure}
\begin{centering}
{\scalebox{0.5}{\includegraphics{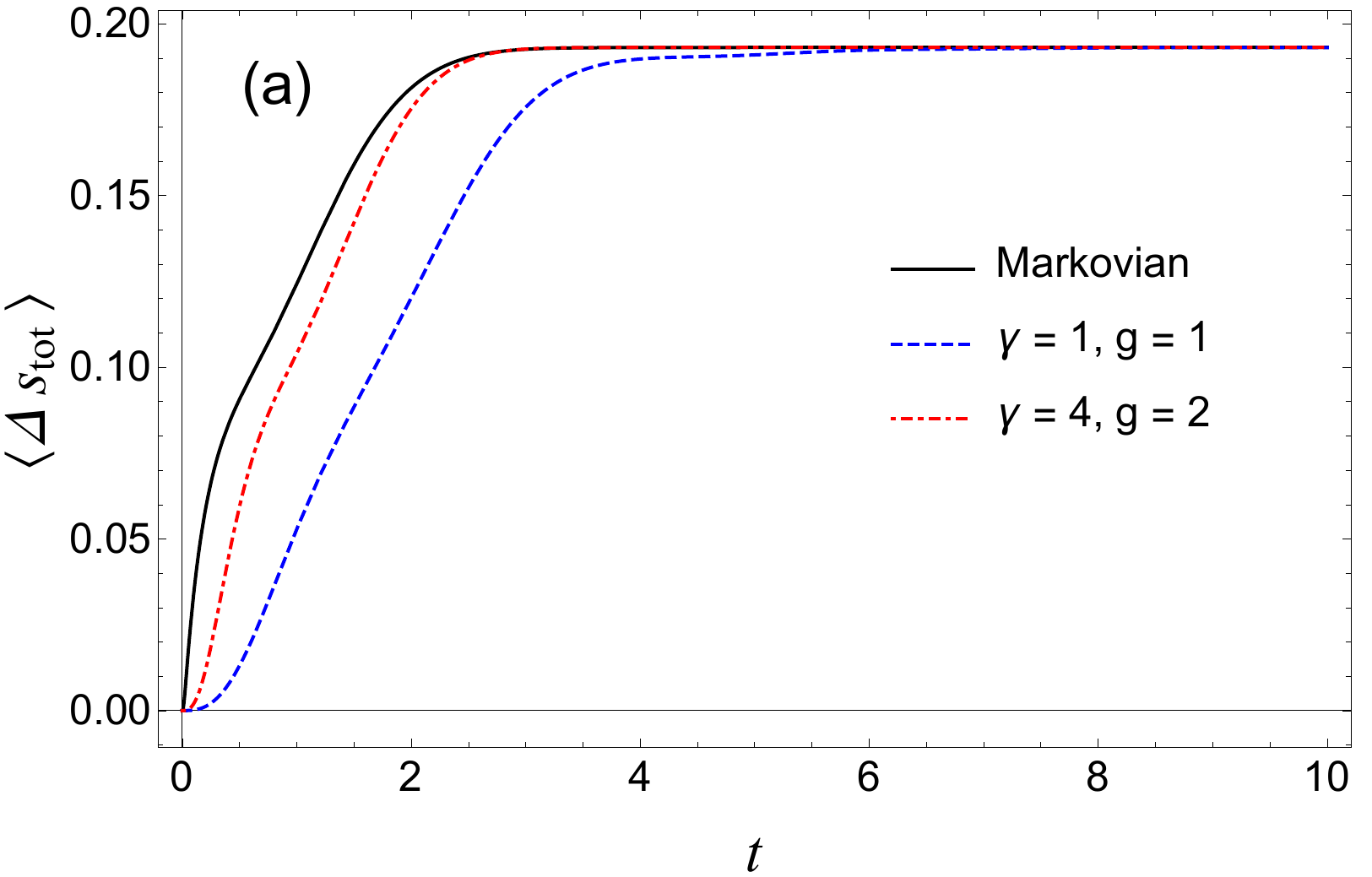}}} {\scalebox{0.35}{\includegraphics{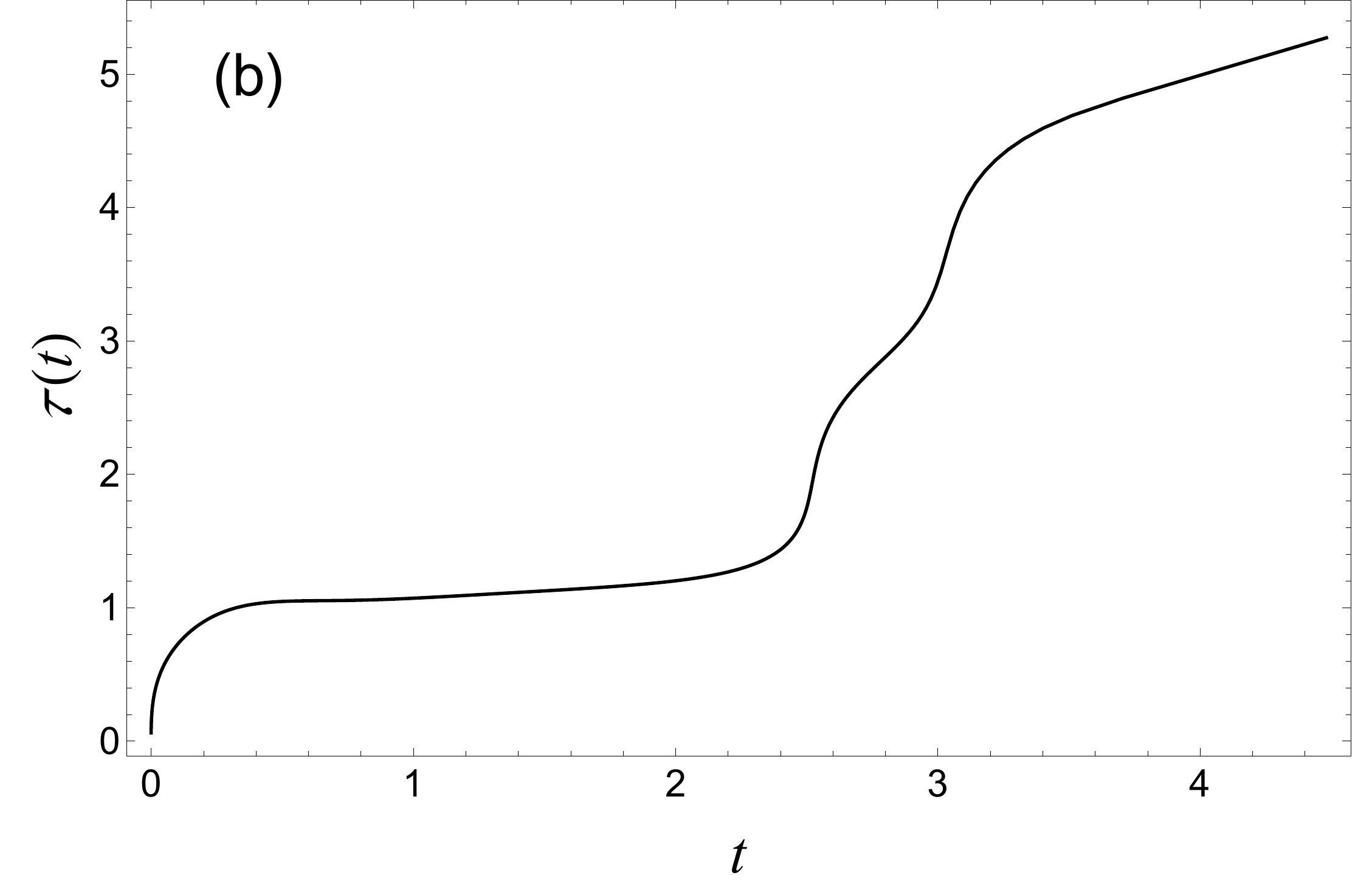}}}
\par\end{centering}
\caption{(a): Evolution of $\langle\Delta s_{\mathrm{tot}}\rangle$ (in units
such that $k_{\mathrm{B}}=1$) for an instantaneous change in environment
temperature from 1 to 2 and various coupling strengths, with the mean
entropy production of the Markovian case plotted for reference. (b):
The delay time between the Markovian ($\tilde{\gamma}=1$) and non-Markovian
($\gamma=1$, $g=1$) cases that produce a given amount of entropy
(see equation (\ref{delayeq})).}
\label{meanentropy}
\end{figure}

The correlations between the non-Markovian environment and the system
mean that the uncertainty in the state of the environment is lower
than that of the equivalent Markovian environment; we can salvage
some information about it from our knowledge of the system's evolution.
Consequently the mean entropy production in the non-Markovian case
will be lower than that of the Markovian case for as long as these
non-equilibrium correlations exist. Dynamically speaking, this means
the non-Markovian mean entropy production should \emph{lag behind}
the Markovian mean entropy production, reaching the same asymptotic
limit but at a later time. Plots of the mean entropy production $\langle\Delta s_{\mathrm{tot}}\rangle$
during the process are shown in Fig. \ref{meanentropy} (a). We can
quantify this delay by comparing the Markovian mean entropy production,
$\langle\Delta s_{\mathrm{tot}}^{\mathrm{M}}\rangle$, with the non-Markovian
mean entropy production $\langle\Delta s_{\mathrm{tot}}^{\mathrm{NM}}\rangle$,
as follows:

\begin{equation}
\langle\Delta s_{\mathrm{tot}}^{\mathrm{M}}\rangle(t)=\langle\Delta s_{\mathrm{tot}}^{\mathrm{NM}}\rangle(t+\tau(t)).\label{delayeq}
\end{equation}

The delay time, $\tau(t)$, illustrated in Fig. \ref{meanentropy}
(b), describes how much later the non-Markovian case will produce
the same amount of entropy, on average, as the Markovian case has
produced at time $t$. The utility of this quantity will become clear
once we discuss the probability density function (pdf) of the entropy
production.

\begin{figure}
\begin{centering}
{\scalebox{0.5}{\includegraphics{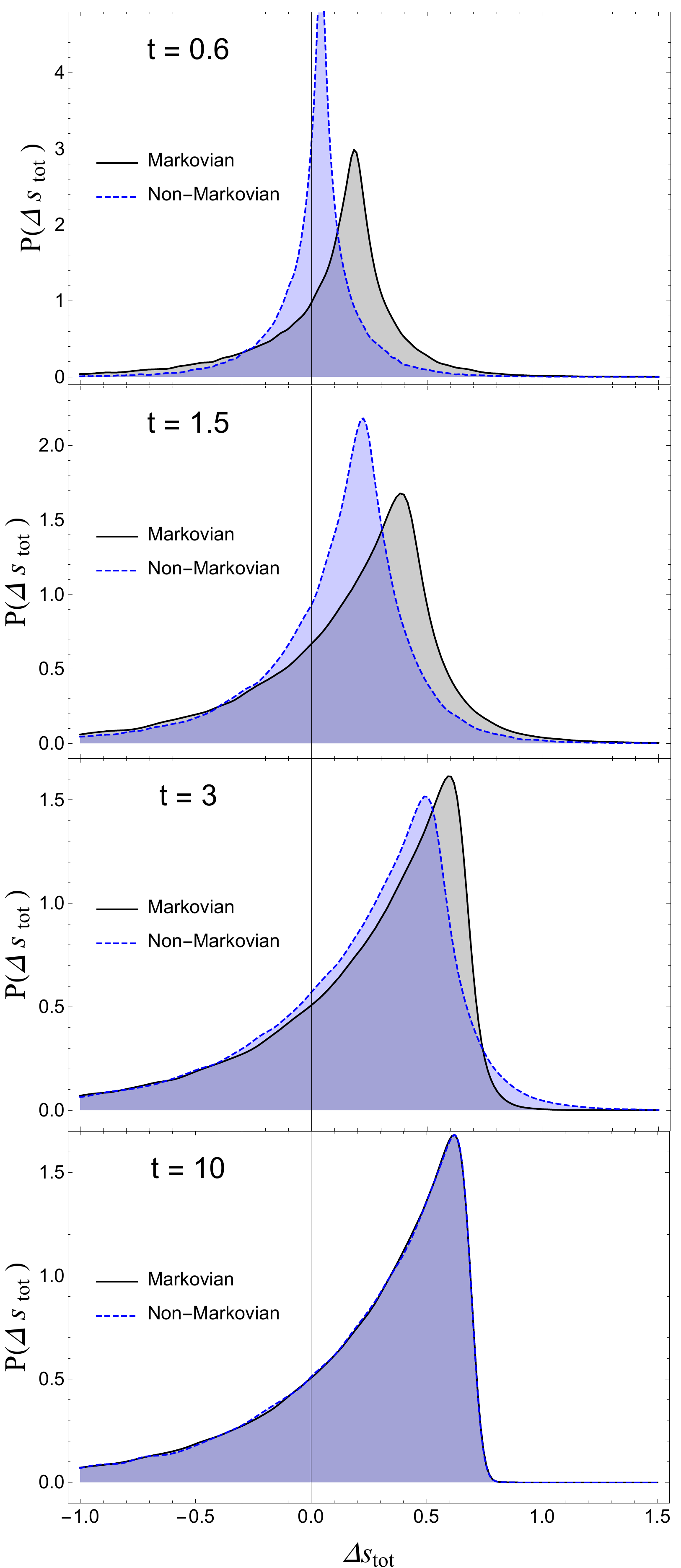}}}
\par\end{centering}
\caption{Evolution of the pdfs of the entropy production (in units such that
$k_{\mathrm{B}}=1$) for an instantaneous change in environment temperature
from 1 to 2, for $g=\gamma=1$, with the pdfs of the corresponding
entropy production for the Markovian case plotted for reference.}
\label{entropypdfs}
\end{figure}

The pdf of entropy production was calculated by simulating 100000
realisations of equations (\ref{tetheredx}) - (\ref{tethereds})
and constructing histograms of $\Delta s_{\mathrm{tot}}$ as a function
of time. This probability density $P(\Delta s_{\mathrm{tot}})$ satisfies
a Detailed Fluctuation Theorem \cite{fluctuationrelations} in the
non-Markovian case automatically because the system and auxiliary
system combined are attached to a Markovian source of noise. The equilibrium
to equilibrium probability density of the entropy production is determined
by the change in $\rho$, which is deterministic, and the probability
density of the heat transfer (see equation (\ref{integratedentropy})).
As discussed, the term $\Delta\ln\rho$ is the same between equilibrium
states for both the Markovian and non-Markovian cases. The distribution
of heat flow \textit{after equilibrium has been reached} is also the
same because the distributions of the variables $x$ and $v$ in the
initial and final equilibrium states are identical. For this reason,
the equilibrium to equilibrium probability densities for the entropy
production should be identical as well. As can be seen in Fig. \ref{entropypdfs}
this is indeed the case. Notice that the entropy production pdfs have
a prominent tail into the negative values. At earlier times, because
of the correlations with the auxiliary system (absent in the Markovian
case), the pdfs necessarily differ.

\begin{figure}[H]
\begin{centering}
{\scalebox{0.5}{\includegraphics{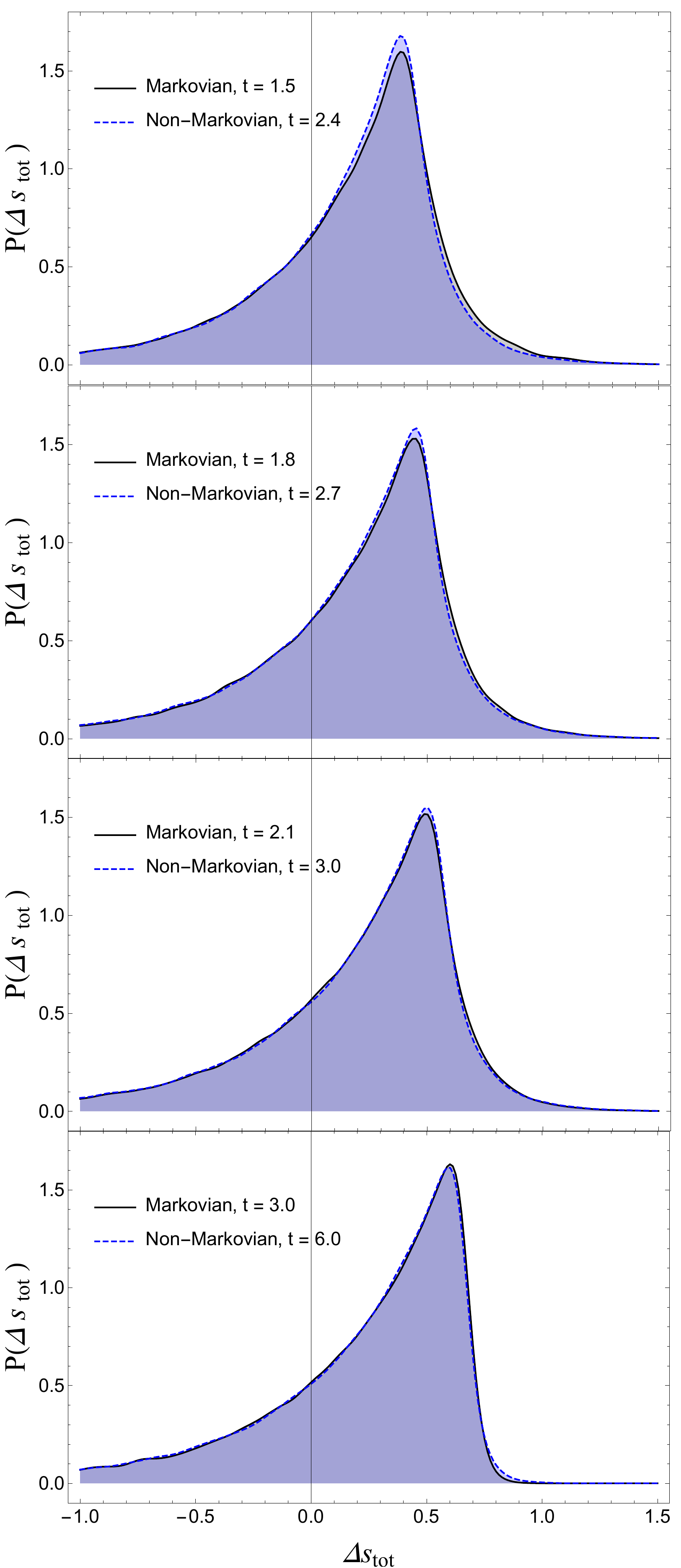}}}
\par\end{centering}
\caption{Comparison of the pdfs of the entropy production of the Markovian
and non-Markovian cases (from Fig. \ref{entropypdfs}, offset by the
delay time shown in Fig. \ref{meanentropy} (b).}
\label{delaypdfs}
\end{figure}

The non-equilibrium pdfs of the dynamical variables of the Markovian
and non-Markovian cases are not directly comparable because of the
correlation between $s$ and $x$ \& $v$. However consideration of
Fig. \ref{delaypdfs} reveals that during relaxation, the pdfs of
Markovian and non-Markovian entropy production adopt the same shapes,
albeit at different times. The difference between these times is exactly
the delay time defined in equation (\ref{delayeq}) and plotted in
Fig. \ref{meanentropy} (b). Therefore the Markovian and non-Markovian
distributions visit the same sequence of shapes, just at different
times.

In summary, the thermodynamics of this type of non-Markovian behaviour
suffer a time delay compared to Markovian thermodynamics. We now discuss
how non-Markovian environments can affect the thermodynamics when
there is a time dependent Hamiltonian.

\section{A Dynamically Tethered Particle\label{sec:A-Dynamically-Tethered}}

\begin{figure}[t]
\begin{centering}
{\scalebox{0.5}{\includegraphics{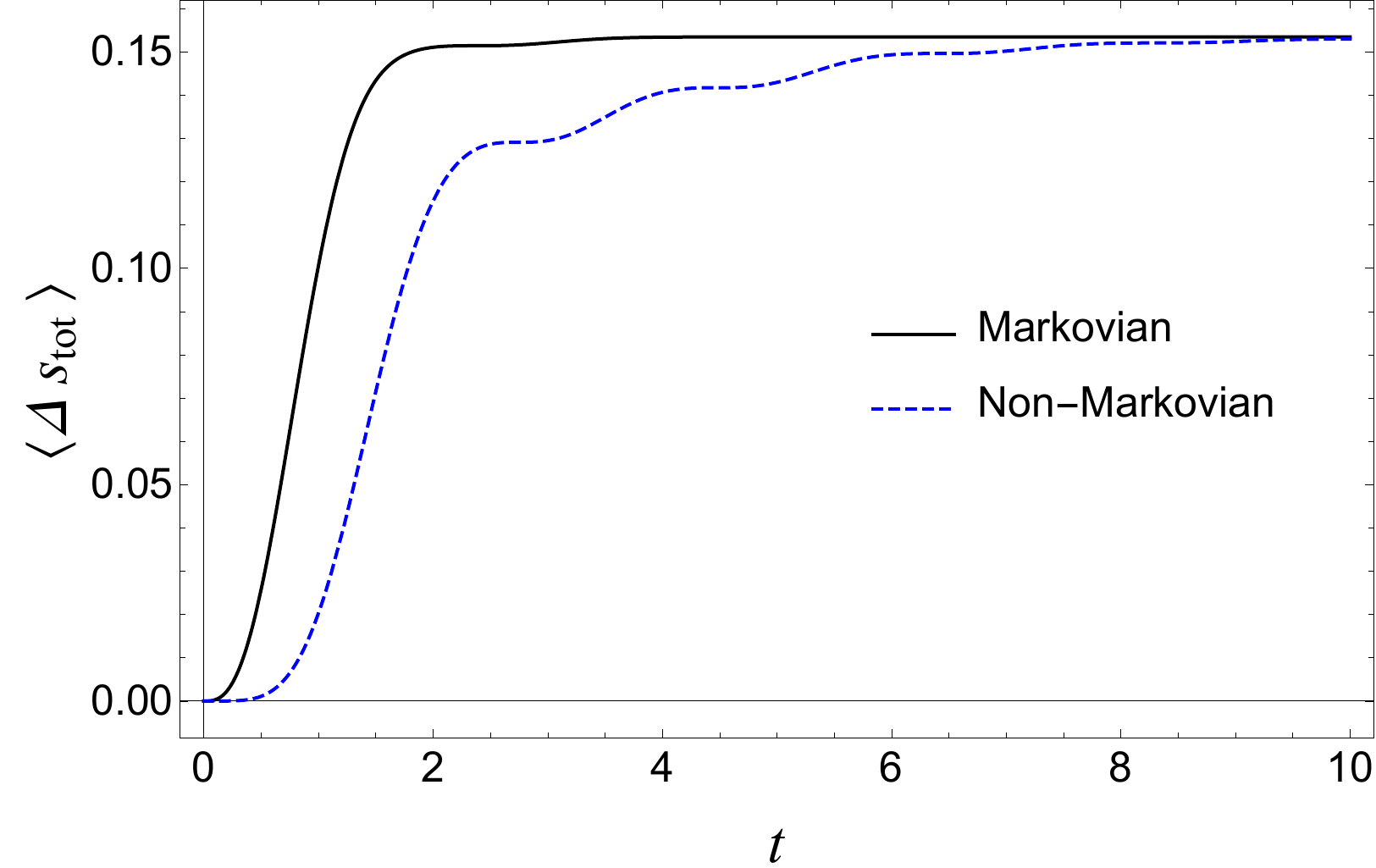}}}
\par\end{centering}
\caption{$\langle\Delta s_{\mathrm{tot}}\rangle$ for an instantaneous change
in $\kappa$ from 1 to 2 and environmental temperature $T=1$. The
non-Markovian case has coupling constants $\gamma=g=1$ and the Markovian
case has coupling constant $\tilde{\gamma}=1$. The asymptotic mean
entropy production is given by equation (\ref{kappaentropy}).}
\label{instentropy}
\end{figure}

The situation we discuss next is a time dependent spring constant
$\kappa$ in the tethered system. The Hamiltonian and equations of
motion take the same form as before except with time-dependent $\kappa(t)$.
The Fokker-Planck equation accommodates $\kappa(t)$ and therefore
our generalised Gaussian form for the non-equilibrium pdf of coordinates
given by equation (\ref{tetheredpdf}) remains employable. The work
done on the system by changing the spring constant is given by

\begin{equation}
\mathcal{W}=\int_{t_{0}}^{t_{1}}\dd t\ \frac{1}{2}\dv{\kappa(t)}{t}x(t)^{2}.\label{workintegralgeneral}
\end{equation}
We consider a linear change in $\kappa$ between $\kappa_{0}$ and
$\kappa_{1}$ over a time period $\Delta t$:

\begin{equation}
\kappa(t)=\begin{cases}
\kappa_{0} & t\leq0\\
\kappa_{0}+(\kappa_{1}-\kappa_{0})\frac{t}{\Delta t} & 0\leq t\leq\Delta t\\
\kappa_{1} & t\ge\Delta t
\end{cases},\label{kappa}
\end{equation}
such that the work done becomes

\begin{equation}
\mathcal{W}=\int_{0}^{\Delta t}\dd t\ \frac{1}{2}\frac{(\kappa_{1}-\kappa_{0})}{\Delta t}x(t)^{2}.\label{workintegral}
\end{equation}
In the limiting case $\Delta t\rightarrow0$ the work done is $\frac{1}{2}(\kappa_{1}-\kappa_{0})x(0)^{2}$
and the Markovian and non-Markovian cases will have identical distributions
of work and entropy production for reasons discussed earlier. The
asymptotic mean entropy production in this case is \cite{fluctuationrelations}
\begin{equation}
\langle\Delta s_{\mathrm{tot}}\rangle=\frac{1}{2}\left(\frac{\kappa_{1}}{\kappa_{0}}-1-\ln\left(\frac{\kappa_{1}}{\kappa_{0}}\right)\right).\label{kappaentropy}
\end{equation}
The time evolution of the mean entropy production for this protocol
is plotted in Fig. \ref{instentropy}. The opposite limit $\Delta t\rightarrow\infty$
is the quasistatic limit. The system and auxiliary system would in
this case never depart from equilibrium with the Markovian bath and
the entropy production vanishes.

If the spring constant is altered more slowly the mean work done on
the system is smaller. The reason for this is as follows: the mean
increment in work is given by $\langle\dd\mathcal{W}\rangle=\frac{1}{2}\langle x^{2}\rangle\dd\kappa$.
If $\dd\kappa$ is positive the effect of coupling to the environment
is for $\langle x^{2}\rangle$ to decrease to match the increased
spring constant. Therefore evolving $\kappa$ more slowly will reduce
the overall mean work done. The argument for the case when $\dd\kappa$
is negative follows in much the same way. The minimum mean work is
therefore performed in the $\Delta t\rightarrow\infty$ limit. Because
of this, we expect the non-Markovian case to see more work performed
on average than the Markovian case when $\kappa$ is continuously
changed since the effects of the bath on the system are delayed (making
the integrand in equation (\ref{workintegralgeneral}) larger for
longer as a result). A higher mean performance of work implies a greater
mean entropy production due to a greater mean transfer of heat into
the bath.

\begin{figure}
\begin{centering}
{\scalebox{0.5}{\includegraphics{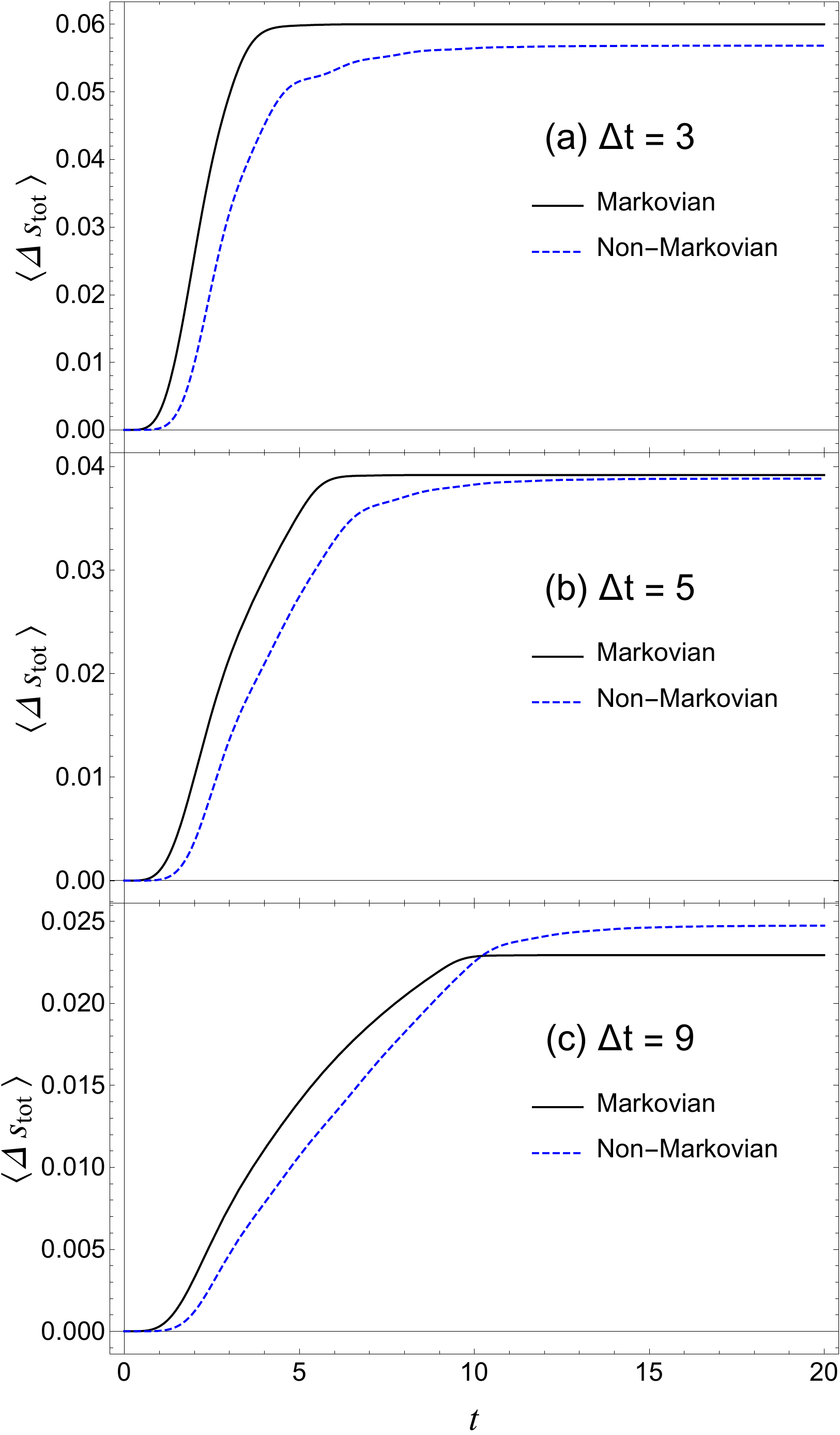}}}
\par\end{centering}
\caption{$\langle\Delta s_{\mathrm{tot}}\rangle$ for a uniform rate of change
of $\kappa$ from 1 to 2 over different time periods $\Delta t$,
with temperature $T=1$. In each case the non-Markovian case has coupling
constants $\gamma=g=1$ and the Markovian case has coupling constant
$\tilde{\gamma}=1$.}
\label{fig:kappaentropy}
\end{figure}

\begin{figure}
\begin{centering}
{\scalebox{0.5}{\includegraphics{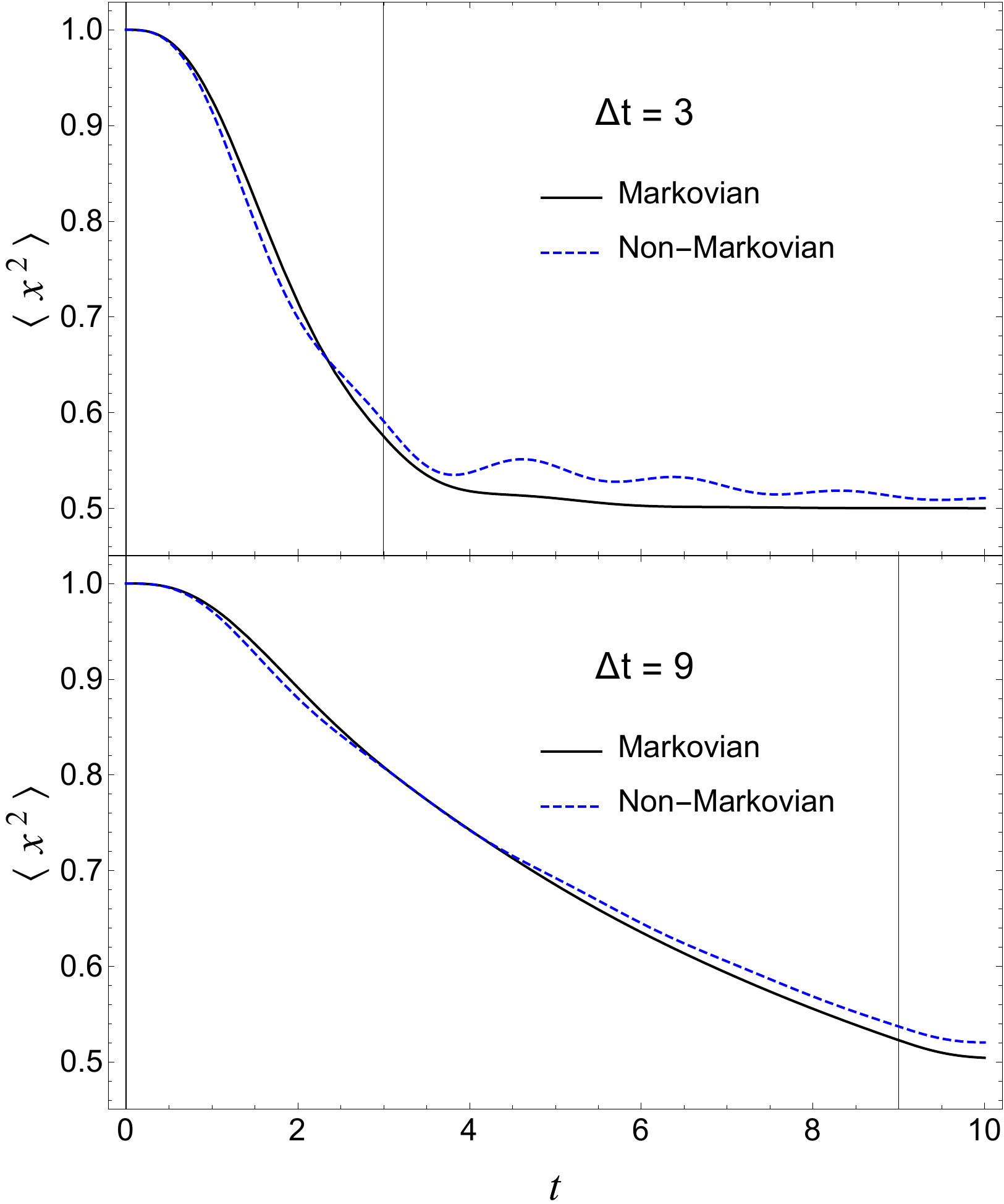}}}
\par\end{centering}
\caption{Evolution of $\langle x^{2}\rangle$ for $g=\gamma=1$ in the non-Markovian
case and $\tilde{\gamma}=1$ in the Markovian case, $T=1$, and for
two different values of $\Delta t$, the duration over which the spring
constant $\kappa$ is changed from 1 to 2.}
\label{kappaxsq}
\end{figure}

The mean entropy production for Markovian and non-Markovian cases
with two protocols of evolution of the spring constant are shown in
Fig. \ref{fig:kappaentropy}. Frame (c) is in accord with the prediction
just described, but in frame (a) the mean non-Markovian entropy production
is lower than that of the Markovian case. Under the logic just proposed
this would mean that the variable $x$ evolves more quickly towards
equilibrium \textit{while the work is being done} in the non-Markovian
case. Fig. \ref{kappaxsq} reveals that this is indeed the case: $\langle x^{2}\rangle$
drops more quickly to begin with, though equilibrium is still reached
more slowly. A Markovian bath will always behave the same way for
a given system - it supplies heat according to its temperature and
does not depend on transient correlations with the system - but depending
on the nature of the protocol it's clear that an auxiliary system
can either delay or quicken a bath's influence.

\section{A Driven Particle\label{sec:A-Driven-System}}

For our final example we consider an untethered particle subject to
a non-conservative oscillatory driving force. With this setup we expect
to be able to produce a non-equilibrium oscillatory steady state accompanied
by a periodic mean entropy production. The It\^{o} processes defining
the dynamics of the non-Markovian case are as follows:

\begin{equation}
\dd v=gs\dd t-\mathcal{F}_{0}\cos(\omega t)\dd t,\label{dfdv}
\end{equation}

\begin{equation}
\dd s=-\gamma s\dd t-gv\dd t+\sqrt{2\gamma T}\dd W.\label{dfds}
\end{equation}
Here $\mathcal{\mathcal{F}}_{0}$ is a force amplitude, $\omega$
the angular frequency, and the driving force performs work on the
system given by 
\begin{equation}
\dd\mathcal{W}=-\mathcal{F}_{0}\cos(\omega t)v\dd t.\label{dfdw}
\end{equation}
The Markovian equivalent case has equation of motion: 
\begin{equation}
\dd v=-\tilde{\gamma}v\dd t-\mathcal{F}_{0}\cos(\omega t)\dd t+\sqrt{2\tilde{\gamma}T}\dd W.\label{mdfdv}
\end{equation}
We assume the following form for the non-Markovian non-equilibrium
pdf: 
\begin{align}
\begin{split}\rho^{\mathrm{NM}}(t,v,s)= & \ \mathrm{exp}\left(N(t)-A(t)v^{2}-B(t)vs\right.\\
 & \left.-C(t)s^{2}-D(t)v-E(t)s\right),
\end{split}
\label{dfpdf}
\end{align}
which in contrast to the previous form contains linear terms in the
exponent. In the Markovian case we use:

\begin{equation}
\rho^{\mathrm{M}}(t,v)=\mathrm{exp}\left(\tilde{N}(t)-\tilde{A}(t)v^{2}-\tilde{B}(t)v\right).\label{mdfpdf}
\end{equation}
The increments in total entropy production are:

\begin{widetext}

\begin{align}
\dd\Delta s_{\mathrm{tot}}^{\mathrm{NM}} & =-\dd t\left(2\gamma-4\gamma TC+\gamma TE^{2}+2\gamma TBEv+4\gamma TCEs+\gamma TB^{2}v^{2}+4\gamma TBCvs\right.\nonumber \\
 & \left.+\left(4\gamma TC^{2}-\frac{\gamma}{T}\right)s^{2}\right)+\dd W\sqrt{2\gamma T}\left(E+Bv+\left(2C-\frac{1}{T}\right)s\right),
\end{align}
and

\begin{align}
\dd\Delta s_{\mathrm{tot}}^{\mathrm{M}} & =-\dd t\left(2\tilde{\gamma}-4\tilde{\gamma}T\tilde{A}+\tilde{\gamma}Tz{\tilde{B}}^{2}+4\tilde{\gamma}T\tilde{A}\tilde{B}v+\left(4\tilde{\gamma}T\tilde{A}^{2}-\frac{\tilde{\gamma}}{T}\right)v^{2}\right)\nonumber \\
 & +\dd W\sqrt{2\tilde{\gamma}T}\left(\tilde{B}+\left(2\tilde{A}-\frac{v}{T}\right)v\right).\label{eq:endfm}
\end{align}

\end{widetext}

\begin{figure}
\begin{centering}
{\scalebox{0.5}{\includegraphics{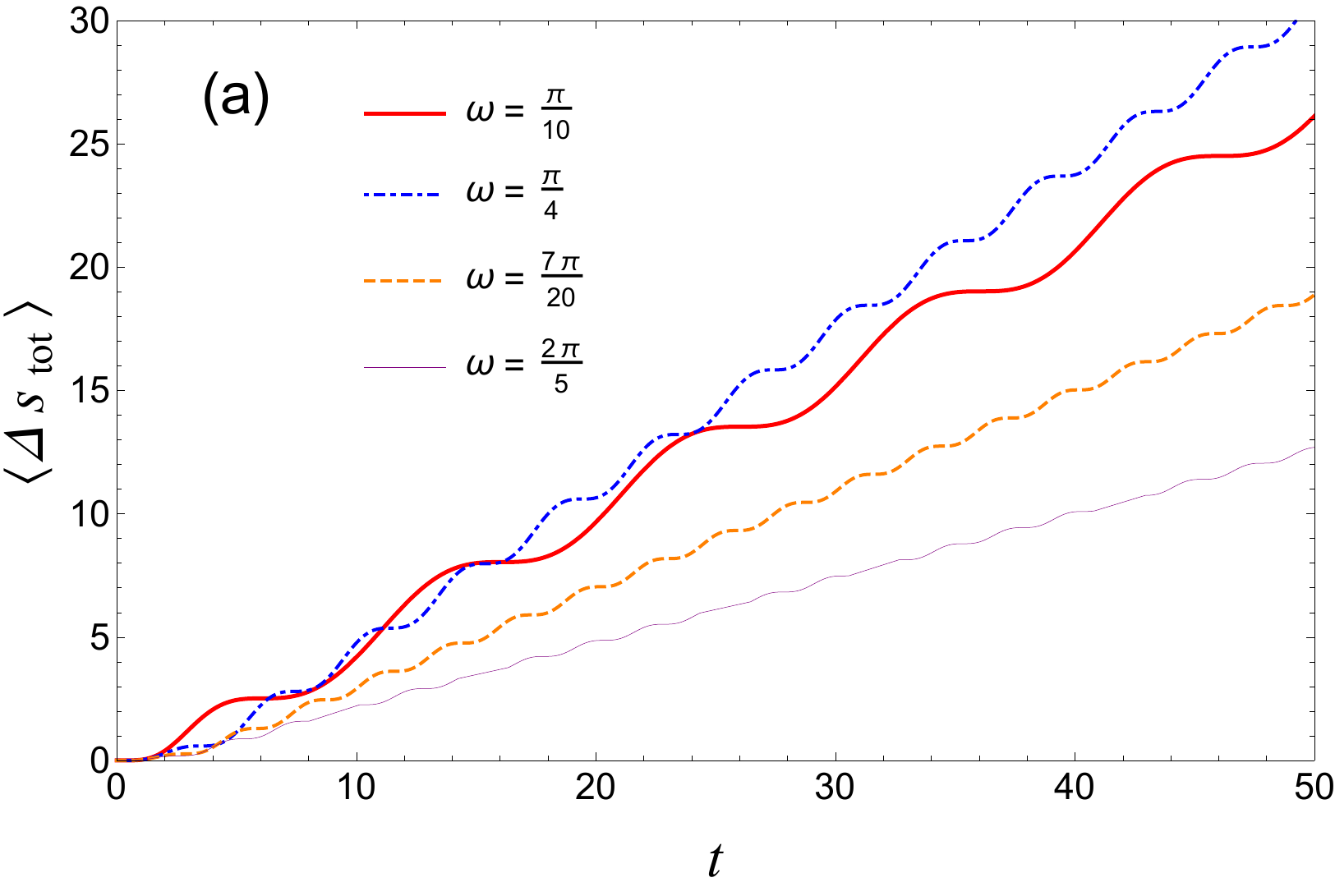}}} {\scalebox{0.5}{\includegraphics{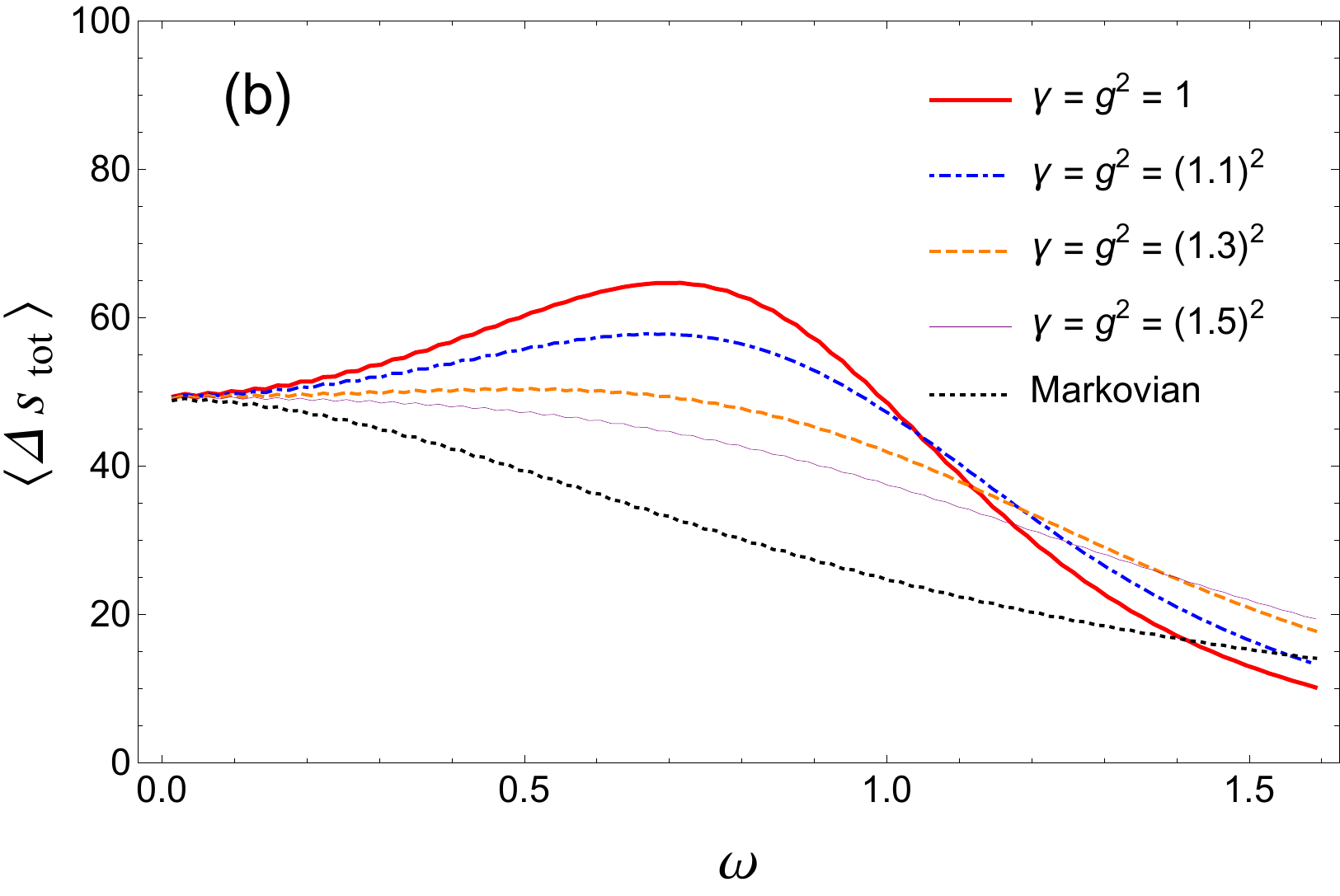}}}
{\scalebox{0.5}{\includegraphics{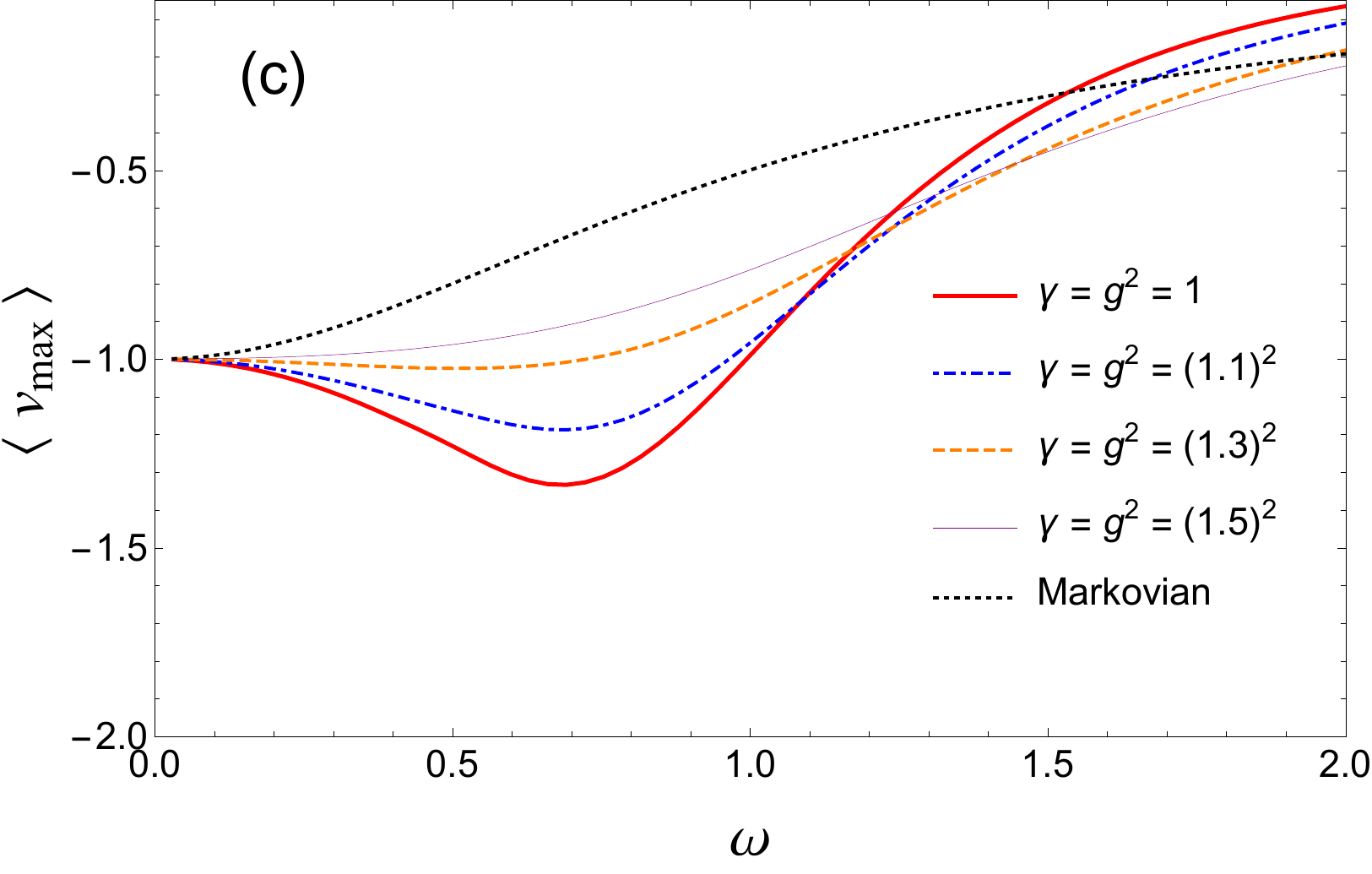}}}
\par\end{centering}
\caption{Evolution of $\langle\Delta s_{\mathrm{tot}}\rangle$ in the periodically
driven system (a) as a function of time in the non-Markovian case
with $g=\gamma=1$ and $T=1$ for different driving frequencies; (b)
as a function of driving frequency at $t=100$ for different coupling
constants, $T=1$, and with the Markovian case at $\gamma=1$ for
reference. (c) Evolution of $\langle v_{\mathrm{max}}\rangle$, the
amplitude of oscillation, as a function of driving frequency at a
given time for different coupling constants, with the Markovian case
at $\gamma=1$ for reference.}
\label{drivensystem}
\end{figure}

After an initial period a non-equilibrium periodic steady state emerges.
Fig. \ref{drivensystem} (a) shows that the time derivative of the
mean total entropy production goes to zero but never becomes negative,
in accordance with the second law of thermodynamics. The mean total
entropy production oscillates in the steady state with frequency equal
to twice the driving frequency because the dissipative term in equation
(\ref{dfds}) produces a positive mean entropy production upon both
the acceleration and deceleration of $s$. Because this dissipative
term $-\gamma s$ (or $-\gamma v$ in the Markovian case) increases
as the amplitude $s$ (or $v$) becomes larger in magnitude, increasing
the driving frequency without also increasing the force will simply
decrease the amplitude and therefore decrease the heat dissipated
into the environment.

Fig. \ref{drivensystem} (b) shows $\langle\Delta s_{\mathrm{tot}}\rangle$
at a given time ($t=100$, which is well within the steady-state regime)
as a function of the driving frequency. The Markovian case agrees
with the prediction just made. However, in the non-Markovian case
the entropy production first increases with frequency before dropping
off as expected. As before, this is a consequence of the correlations
between the auxiliary system that cause the external protocol (in
this case a non-conservative driving force rather than a change of
spring constant) to have heightened effects (see the amplitude of
oscillation in $\langle v\rangle$ as a function of driving frequency
in Fig. \ref{drivensystem} (c)). 

\section{Conclusions\label{sec:Conclusions}}

In summary, we have performed an extension of stochastic thermodynamics
to environments which can correlate with a system of interest. This
was done by interposing an auxiliary system between the system of
interest and a heat bath. The system thereby feels the influence of
non-Markovian noise while the results of stochastic thermodynamics
automatically apply due to the fact that the system plus auxiliary
system are still coupled to a Markovian bath.

The presence of memory can heighten or dampen the influence of the
environment, depending on the time scales involved. When there is
no time-dependent external work protocol and the system merely relaxes,
the presence of memory simply serves to delay the equilibration: all
distributions of stochastic entropy production lie on the same continuum
of distributions exhibited by the equivalent Markovian system. When
the external protocol is time-dependent, the combination of non-Markovian
thermal coupling and the mechanical coupling can allow the work and
entropy production to either exceed or be exceeded by those of the
equivalent Markovian system. Because the Markovian approximation is
an idealised limit, these observations shed light into how the thermodynamics
might differ from this idealisation in systems where time scales aren't
so easily stratified.

The implementation of unphysical degrees of freedom to reproduce the
desired memory effects circumvents the difficulty in both the theoretical
and computational analysis of memory kernels. The results shown in
this work serve as a stimulus for future studies of non-Markovian
stochastic thermodynamics using the auxiliary system framework.

\appendix

\section{Solution to Fokker-Planck equations and Derivation of entropy production}

The Fokker-Planck equation for the harmonically tethered system coupled
to a non-Markovian environment, described in equations (\ref{tetheredx})
- (\ref{tethereds}), is

\begin{align}
\begin{split}\pdv{\rho^{\mathrm{NM}}}{t}= & -\pdv{(v\rho^{\mathrm{NM}})}{x}-\pdv{((-\kappa x+gs)\rho^{\mathrm{NM}})}{v}\\
 & +\pdv{((\gamma s+gv)\rho^{\mathrm{NM}})}{s}+\gamma T\pdv[2]{\rho^{\mathrm{NM}}}{t}.
\end{split}
\end{align}
Evaluating each of these terms using the trial solution in equation
(\ref{tetheredpdf}) we get:

\begin{equation}
\pdv{\rho^{\mathrm{NM}}}{t}=\left(\dot{N}-\dot{A}x^{2}-\dot{B}v^{2}-\dot{C}s^{2}-\dot{D}xv-\dot{E}xs-\dot{F}vs\right)\rho^{\mathrm{NM}},
\end{equation}

\begin{equation}
-\pdv{(v\rho^{\mathrm{NM}})}{x}=-v(-2Ax-Dv-Es)\rho^{\mathrm{NM}},
\end{equation}

\begin{equation}
\pdv{(\kappa x\rho^{\mathrm{NM}})}{v}=\kappa x(-2Bv-Dx-Fs)\rho^{\mathrm{NM}},
\end{equation}

\begin{equation}
-\pdv{(gs\rho^{\mathrm{NM}})}{v}=-gs(-2Bv-Dx-Fs)\rho^{\mathrm{NM}},
\end{equation}

\begin{equation}
\pdv{(\gamma s\rho^{\mathrm{NM}})}{s}=\gamma\rho^{\mathrm{NM}}+\gamma s(-2Cs-Ex-Fv)\rho^{\mathrm{NM}},
\end{equation}

\begin{equation}
\pdv{(gv\rho^{\mathrm{NM}})}{s}=gv(-2Cs-Ex-Fv)\rho^{\mathrm{NM}},
\end{equation}

\begin{equation}
\gamma T\pdv[2]{\rho^{\mathrm{NM}}}{s}=\gamma T(-2Cs-Ex-Fv)^{2}\rho^{\mathrm{NM}}+\gamma T(-2C)\rho^{\mathrm{NM}}.
\end{equation}
From here we can read off the coefficients to $\rho^{\mathrm{NM}}$,
$x^{2}\rho^{\mathrm{NM}}$, $xv\rho^{\mathrm{NM}}$ etc to write down
the evolution equations for $N$ and $A-F$:

\begin{equation}
\dot{N}=\gamma-2\gamma TC,\label{ndot}
\end{equation}

\begin{equation}
-\dot{A}=-\kappa D+\gamma TE^{2},
\end{equation}

\begin{equation}
-\dot{B}=D-gF+\gamma TF^{2},
\end{equation}

\begin{equation}
-\dot{C}=gF-2\gamma C+4\gamma TC^{2},
\end{equation}

\begin{equation}
-\dot{D}=2A-2\kappa B-gE+2\gamma TEF,
\end{equation}

\begin{equation}
-\dot{E}=-\kappa F+gD-\gamma E+4\gamma TCE,
\end{equation}

\begin{equation}
-\dot{F}=E+2gB-gE-\gamma F-2gC+4\gamma TCF.\label{fdot}
\end{equation}
The total entropy production is written $\dd\Delta s_{\mathrm{tot}}^{{\rm {NM}}}=\dd\Delta s_{\mathrm{sys}}^{{\rm {NM}}}+\dd\Delta s_{\mathrm{bath}}^{{\rm {NM}}}$.
We compute the first contribution from the pdf over dynamical variables:
\begin{align}
\begin{split}\dd\Delta s_{\mathrm{sys}}^{{\rm {NM}}} & =-\dd\ln\rho^{\mathrm{NM}}\\
 & =-\pdv{\ln\rho^{\mathrm{NM}}}{t}\dd t-\pdv{\ln\rho^{\mathrm{NM}}}{x}\dd x-\pdv{\ln\rho^{\mathrm{NM}}}{v}\dd v\\
 & -\pdv{\ln\rho^{\mathrm{NM}}}{s}\dd s-\gamma T\pdv[2]{\ln\rho^{\mathrm{NM}}}{s}\dd t,
\end{split}
\label{dss}
\end{align}
where we've used It\^{o}'s lemma. Again, we compute each term:

\begin{equation}
-\pdv{\ln\rho^{\mathrm{NM}}}{t}\dd t=-\dot{N}+\dot{A}x^{2}+\dot{B}v^{2}+\dot{C}s^{2}+\dot{D}xv+\dot{E}xs+\dot{F}vs,\label{dsdecomp1}
\end{equation}

\begin{equation}
-\pdv{\ln\rho^{\mathrm{NM}}}{x}\dd x=(2Ax+Dv+Es)(v\dd t),
\end{equation}

\begin{equation}
-\pdv{\ln\rho^{\mathrm{NM}}}{v}\dd v=(2Bv+Dx+Fs)(-\kappa x\dd t++gs\dd t),
\end{equation}

\begin{equation}
-\pdv{\ln\rho^{\mathrm{NM}}}{s}\dd s=(2Cs+Ex+Fs)(-\gamma s\dd t-gv\dd t+\sqrt{2\gamma T}\dd W),
\end{equation}

\begin{equation}
-\gamma T\pdv[2]{\ln\rho^{\mathrm{NM}}}{s}=2\gamma TC\dd t.\label{dsdecomp2}
\end{equation}
The entropy production in the bath is given by $\dd Q / T$, where $Q$ is the heat flow into the bath from the auxiliary system. The energy of the auxiliary system can change either by this heat flow or by doing work on the system. To identify these components, we examine the energy differential of the auxiliary system (using It\^{o}'s lemma):
\begin{align}
\begin{split}
    \frac{1}{2} \dd \left( s^2 \right) =& \frac{1}{2} \left (2 s \dd s + 2 \gamma T \dd t \right) \\
    =& \frac{1}{2} \Big( -2 \gamma s^2 \dd t - 2 g v s \dd t + 2 s \sqrt{2 \gamma T} \dd W  \\
    &  + 2 \gamma T \dd t \Big) \\
    =& \left(-\gamma s^2 - g v s + \gamma T \right) \dd t + s \sqrt{2 \gamma T} \dd W.
\end{split}
\label{energysplit}
\end{align}
By inspection of the system's kinetic energy, we can identify the $g v s \dd t$ as the differential of the work done by the coupling between the system and auxiliary system:
\begin{align}
\begin{split}
    \frac{1}{2} \dd \left( v^2 \right) =& 2 v \dd v \\
    =& - \kappa x v \dd t + g v s \dd t.
\end{split}
\label{energysplit2}
\end{align}
The remaining part of Eq. \ref{energysplit} is therefore the heat flow to the bath, which allows us to identify the entropy production in the bath:
\begin{align}
\begin{split}\dd\Delta s_{\mathrm{bath}}^{{\rm {NM}}} & =-\frac{\dd Q}{T},\\
 & =-\frac{\dd(s^{2})}{T} - \frac{1}{T} g v s \dd t,\\
 & =\dd t\left(\frac{\gamma s^{2}}{T}-\gamma\right)+\dd W\left(\frac{s}{T}\sqrt{2\gamma T}\right),
\end{split}
\label{dse}
\end{align}
again using It\^{o}'s lemma. The same expression can be derived using considerations laid out in \cite{spinneyfordpre}. Then we substitute equations (\ref{dsdecomp1}
- \ref{dsdecomp2}) into equation (\ref{dss}), add that to (\ref{dse}),
and then compare coefficients of terms in $\dd W$, $\dd t$, $x^{2}\dd t$,
$xv\dd t$ etc. Most of the terms in equations (\ref{ndot} - \ref{fdot})
cancel, and the final result is equation (\ref{tetheredentropy}).

The Fokker-Planck equation for the Markovian dynamical system described
in equations (\ref{markoviantetheredx}) - (\ref{markoviantetheredv})
is

\begin{align}
\begin{split}\pdv{\rho^{\mathrm{M}}}{t}= & -\pdv{(v\rho^{\mathrm{M}})}{x}-\pdv{((-\kappa x-\tilde{\gamma}v)\rho^{\mathrm{M}})}{v}\\
 & +\tilde{\gamma}T\pdv[2]{\rho^{\mathrm{M}}}{t},
\end{split}
\end{align}
and as for the non-Markovian case we write:

\begin{equation}
\pdv{\rho^{\mathrm{M}}}{t}=\left(\dot{\tilde{N}}-\dot{\tilde{A}}x^{2}-\dot{\tilde{B}}v^{2}-\dot{\tilde{C}}xv\right)\rho^{\mathrm{M}}
\end{equation}

\begin{equation}
-\pdv{(v\rho^{\mathrm{M}})}{x}=-v(-2\tilde{A}x-\tilde{C}v)\rho^{\mathrm{M}},
\end{equation}

\begin{equation}
\pdv{(\kappa x\rho^{\mathrm{M}})}{v}=\kappa x(-2\tilde{B}v-\tilde{C}x)\rho^{\mathrm{M}},
\end{equation}

\begin{equation}
\pdv{(\tilde{\gamma}v\rho^{\mathrm{M}})}{v}=\tilde{\gamma}\rho^{\mathrm{M}}+\tilde{\gamma}v(-2\tilde{B}v-\tilde{C}x)\rho^{\mathrm{M}},
\end{equation}

\begin{equation}
\tilde{\gamma}T\pdv[2]{\rho^{\mathrm{M}}}{v}=\tilde{\gamma}T(-2\tilde{B}v-\tilde{C}x)^{2}\rho^{\mathrm{M}}+\tilde{\gamma}T(-2\tilde{B})\rho^{\mathrm{M}}.
\end{equation}
Again we read off the coefficients to write down the evolution equations:

\begin{equation}
\dot{\tilde{N}}=\tilde{\gamma}-2\tilde{\gamma}T\tilde{B},\label{markovianndot}
\end{equation}

\begin{equation}
-\dot{\tilde{A}}=-\kappa\tilde{C}+\tilde{\gamma}T\tilde{C}^{2},
\end{equation}

\begin{equation}
-\dot{\tilde{B}}=\tilde{C}-2\tilde{\gamma}\tilde{B}+4\tilde{\gamma}T\tilde{B}^{2},
\end{equation}

\begin{equation}
-\dot{\tilde{C}}=2\tilde{A}-2\kappa\tilde{B}-\tilde{\gamma}\tilde{C}+4\tilde{\gamma}T\tilde{B}\tilde{C}.\label{markoviancdot}
\end{equation}
We substitute these into the system entropy production: 
\begin{align}
\begin{split}\dd\Delta s_{\mathrm{sys}}^{{\rm {M}}} & =-\dd\ln\rho^{\mathrm{M}}\\
 & =-\pdv{\ln\rho^{\mathrm{M}}}{t}\dd t-\pdv{\ln\rho^{\mathrm{M}}}{x}\dd x-\pdv{\ln\rho^{\mathrm{M}}}{v}\dd v\\
 & -\tilde{\gamma}T\pdv[2]{\ln\rho^{\mathrm{M}}}{v}\dd t,
\end{split}
\label{markoviandss}
\end{align}
where we've used It\^{o}'s lemma. Again, we decompose each term.

\begin{equation}
-\pdv{\ln\rho^{\mathrm{M}}}{t}\dd t=-\dot{\tilde{N}}+\dot{\tilde{A}}x^{2}+\dot{\tilde{B}}v^{2}+\dot{\tilde{C}}xv,\label{markoviandsdecomp1}
\end{equation}

\begin{equation}
-\pdv{\ln\rho^{\mathrm{M}}}{x}\dd x=(2\tilde{A}x+\tilde{C}v)(v\dd t),
\end{equation}

\begin{equation}
-\pdv{\ln\rho^{\mathrm{M}}}{v}\dd v=(2\tilde{B}v+\tilde{C}x)(-\kappa x\dd t-\tilde{\gamma}v\dd t+\sqrt{2\tilde{\gamma}T}\dd W),
\end{equation}

\begin{equation}
-\tilde{\gamma}T\pdv[2]{\ln\rho^{\mathrm{M}}}{s}=2\tilde{\gamma}TC\dd t.\label{markoviandsdecomp2}
\end{equation}
The bath term is

\begin{align}
\begin{split}\dd\Delta s_{\mathrm{bath}}^{{\rm {M}}} & =-\frac{\dd Q}{T},\\
 & =-\frac{\dd(v^{2})}{T} - \kappa x v \frac{1}{T} \dd t,\\
 & =-\frac{2v\dd v}{T}-\tilde{\gamma}\dd t,\\
 & =\dd t\left(\frac{\tilde{\gamma}v^{2}}{T}-\tilde{\gamma}\right)+\dd W\left(\frac{v}{T}\sqrt{2\tilde{\gamma}T}\right).
\end{split}
\label{markoviandse}
\end{align}
Proceeding in the same way as the non-Markovian case, substitution
and cancellation results in equation (\ref{markoviantetheredentropy}).

Similar procedures are followed to solve the appropriate Fokker-Planck
equations and derive entropy productions in the other example cases.
\end{document}